\documentstyle[a4,12pt]{article}
\begin{document}
\begin{titlepage} \vspace{0.2in} \begin{flushright}
MITH-98/2 \\ \end{flushright} \vspace*{1.5cm}
\baselineskip=30pt
\begin{center} {\LARGE \bf Comment on "Anomalous Deep Inelastic 
Scattering from Liquid H$_2$O-D$_2$O: Evidence of Nuclear Quantum 
Entanglement"}
\end{center}
\baselineskip=12pt
\vspace*{0.8cm}
\begin{center}
{\bf M.~Buzzacchi, E.~Del~Giudice, G.~Preparata}\\ \vspace*{0.5cm}
Dept.~of~Physics,~Universit\`{a}~Statale~di~Milano\\
and INFN - Istituto Nazionale di Fisica Nucleare\\
Via~Celoria~16,~20133 Milano,~Italy\\
\end{center}
 
\baselineskip=12pt 
\vfill \begin{flushleft}  20 January 1998\\
\end{flushleft}
\end{titlepage}
\baselineskip=24pt

In a recent Letter \cite{1}, C.A.~Chatzidimitrou-Dreismann {\it et 
al.} report on a very interesting series of experiments of 
deep-inelastic scattering off H$_2$O, D$_2$O and H$_2$O-D$_2$O 
mixtures with different $x_D$, the D$_2$O molar fraction.

While we agree with the authors that their results are in dramatic 
disagreement with the very reasonable expectation that the 
cross-section ratio $Q=\frac{\sigma_{H}}{\sigma_{D}}$ {\it should be 
independent of} $x_D$ (see Fig.~2 of their paper), we strongly 
disagree with their conclusion that Quantum Entanglement (QE) is a 
possible explanation of their remarkable observations.~~~Indeed, 
whereas for the analogous results obtained in Raman scattering 
\cite{2} QE remains a logical, though quite arduous, possibility, due 
to the non-local nature of the interaction of the laser field with the 
water molecules, for the deep inelastic scattering process analysed in 
Ref.~[1] the high values of the momentum transfers (or, equivalently, 
the small values, $< 0.1 \AA$, of the wave-lengths involved) make any 
spatial coherence among different nuclei essentially impossible.~~~Let 
us recall that it is precisely the basic incoherence of the scattering 
nuclei (H, D~or~O) that lies at the roots of the validity of the 
"impulse approximation", which appears to give an adequate description 
of the experimental data~[1].

If not QE, what else can then explain the observations of Refs.~[1] 
and [2]~?~~~We would like to suggest that the surprising 
$x_D$-dependencies found in the latter papers can be easily and 
naturally understood within a new approach to condensed matter 
described in a recent book \cite{3}, and applied to water in 
Ref.~\cite{4}.~~~According to QED-coherence (for in this approach one 
takes into account the many-body coherent electrodynamic interaction 
among the water molecules) water consists of two interpenetrating 
fluids, one consisting of molecules oscillating in phase with a 
classical electromagnetic field, while the other comprises a dense 
vapour of incoherent, independent molecules.   It is now perfectly 
reasonable that deep inelastic neutron scattering turns out to be 
different on water molecules belonging to the different fluids, and 
that the incoherent fluid scatters the neutrons more strongly 
than the coherent one (which can transfer quite a lot of momentum to 
the center of mass, like it happens in the M\"{o}ssbauer effect): in 
this way the observed cross sections are given by:
\begin{equation}
\sigma_{H,D}=\sigma_{H,D}^{(i)}\left(\epsilon_{H,D}\frac{N_{H,D}^{(c)}}
{N_{H,D}}+
\frac{N_{H,D}^{(i)}}{N_{H,D}}\right),
\end{equation}
where $N_{H,D}^{(i)}$, $N_{H,D}^{(c)}$ and 
$N_{H,D}=N_{H,D}^{(i)}+N_{H,D}^{(c)}$ are the numbers of incoherent, 
coherent and total H$_2$O and D$_2$O molecules respectively, 
$\sigma_{H,D}^{(i)}$ the deep-inelastic cross section of the 
incoherent fluids and $\epsilon_{H,D}<1$ the ratio between the 
coherent and the incoherent cross-section.

Due to the two-fluid nature of both H$_2$O and D$_2$O, the fraction 
$\xi_{H,D}=\frac{N_{H,D}^{(i)}}{N_{H,D}}$ of incoherent molecules, 
which in the unmixed liquids at $T=300~K$ 
are  $\xi_{H}\cong\xi_{D}\cong0.7$~~~[4], will in general depend on the 
molar fraction $x_D$, for thermodynamic equilibrium in the incoherent 
"vapour-like" phase requires:
\begin{equation}
\frac{N_{D}^{(i)}}{N_{D}}\rightarrow 1~~~~~~~  
(x_{D}\rightarrow 0),
\end{equation}
\begin{equation}
\frac{N_{H}^{(i)}}{N_{H}}\rightarrow 1~~~~~~~ 
(x_{D}\rightarrow 1),
\end{equation}
due to the decrease of the chemical potential of the incoherent 
molecules of the highly diluted species. As a result, calling 
$Q^{(0)}=\left(\frac{\sigma_{H}}{\sigma_{D}}\right)_{pure}$ the ratio 
of the cross sections in the 
pure phases:
\begin{equation}
Q^{(0)}=\frac{\sigma_{H}^{(i)}}{\sigma_{D}^{(i)}}\frac{\left[\epsilon_{H}\left(
1-\xi_{H}\right)+\xi_{H}\right]}{\left[\epsilon_{D}\left(
1-\xi_{D}\right)+\xi_{D}\right]}
\end{equation}
one obtains the limits:
\begin{equation}
\frac{\sigma_{H}}{\sigma_{D}}\rightarrow Q^{(0)}
\left[\epsilon_{D}\left(
1-\xi_{D}\right)+\xi_{D}\right]~~~~~(x_{D}\rightarrow 0),
\end{equation}
\begin{equation}
\frac{\sigma_{H}}{\sigma_{D}}\rightarrow Q^{(0)}\frac{1}
{\left[\epsilon_{H}\left(
1-\xi_{H}\right)+\xi_{H}\right]}~~~~~(x_{D}\rightarrow 1),
\end{equation}
which for $\xi_{D}\cong\xi_{H}\cong0.7$ and $\epsilon_{D}\cong0$, 
$\epsilon_{H}\cong0.5$ gives an adequate representation of the 
experimental data. A similar analysis, which shall be reported 
elsewhere \cite{5}, gives 
good account of the other reported anomalies of H$_2$O-D$_2$O mixtures 
for Raman scattering [2] and H$^+$,~D$^+$ conductances \cite{6}.

In conclusion, we would like to emphasize that the surprising 
anomalies that have been revealed in H$_{2}$O-D$_{2}$O mixtures 
instead of the problematic (better, untenable) QE appear to give further 
support to the two-fluid structure of water and to the theory of QED 
coherence~[3] that unambiguously predicts it.

\end{document}